\documentclass[a4paper]{jpconf}
\usepackage{graphicx}
\begin{document}
\title{Bound-state properties from field-theory correlators}
\author{Dmitri Melikhov}
\address{
HEPHY, Austrian Academy of Sciences, Nikolsdorfergasse 18, A-1050 Vienna, Austria\\ 
Faculty of Physics, University of Vienna, Boltzmanngasse 5, A-1090 Vienna, Austria\\ 
SINP, Moscow State University, 119991, Moscow, Russia}
\ead{dmitri\_melikhov@gmx.de}
\begin{abstract}
We discuss the details of calculating hadron properties from the OPE for correlators of quark currents in QCD, 
which constitutes the basis of the method of QCD sum rules. The main emphasis
is laid on gaining control over the systematic uncertainties of the hadron parameters obtained
within this method. We start with examples from quantum mechanics, where bound-state properties may be 
calculated independently in two ways: exactly, by solving the Schr\"odinger equation,
and approximately, by the method of sum rules. Knowing the exact solution allows us to control
each step of the sum-rule extraction procedure. On the basis of this analysis, we formulate several
improvements of the method of sum rules. We then apply these modifications to the analysis of
the decay constants of heavy mesons.
\end{abstract}

\section{Introduction}
The method of dispersive sum rules \cite{svz} is one of the widely used methods for 
obtaining properties of ground-state hadrons in QCD. The method involves two steps: 
(i) one calculates the relevant correlator in QCD at relatively small values of the Eucledian time;  
(ii) one applies numerical procedures suggested by quark-hadron duality in order to isolate the ground-state 
contribution from this correlator. These numerical procedures cannot determine a single value of the 
ground-state parameter but should provide the band of values containing the true 
hadron parameter. This band is a systematic, or intrinsic, uncertainty of the method of sum rules. 
 
An unbiased judgement of the reliability of the extraction procedures adopted in the method of sum rules 
may be acquired by applying these procedures to problems where the ground-state parameters
may be found independently and exactly as soon as the parameters 
of theory are fixed. Presently, only quantum-mechanical potential models 
provide such a possibility: 
(i) the bound-state parameters (masses, wave functions, form factors) are known precisely from the Schr\"odinger equation;
(ii) direct analogues of the QCD correlators may be calculated exactly. 

Making use of these models, we studied the extraction of ground-state parameters 
from different types of correlators: namely, the ground-state decay 
constant from two-point vacuum-to-vacuum correlator \cite{lms_2ptsr}, the form factor from three-point 
vacuum-to-vacuum correlator \cite{lms_3ptsr}, and the form factor from vacuum-to-hadron correlator \cite{m_lcsr}. 
We have demonstrated that the standard procedures adopted in the method of sum rules not always work properly: 
the true value of the bound-state parameter was shown to lie outside 
the band obtained according to the standard criteria. 
These results gave us a solid ground to claim that also in QCD the actual accuracy of the method may be 
worse than the accuracy expected on the basis of applying the standard criteria. 

We realized that the main origin of these problems of the method originate from an over-simplified model 
for hadron continuum which is described as a perturbative contribution above a constant Borel-parameter 
independent effective continuum threshold. 
We introduced the notion of the {\it exact} effective continuum threshold, 
which corresponds to the true bound-state parameters: in potential models the true hadron 
parameters---decay constant and form factor---are known and the exact effective continuum 
thresholds for different correlators may be calculated. 
We have demonstrated that the exact effective continuum threshold 
(i) is not a universal quantity and depends on the correlator considered (i.e., it is in general 
different for two-point and three-point vacuum-to-vacuum correlators), 
(ii) depends on the Borel parameter and, for the form-factor case, also on the momentum transfer \cite{lms_3ptsr,m_lcsr,irina}. 

In recent publications \cite{lms_new} we proposed a new algorithm for extracting 
the parameters of the ground state. 
The idea formulated in these papers is to relax the standard assumption of a Borel-parameter 
independent effective continuum threshold and to allow for a Borel-parameter dependent quantity. 
This lecture explains the details of this procedure and its application to decay constants of heavy mesons.


\section{\label{Sect:QM}OPE and sum rule in quantum-mechanical potential model}
Let us start with quantum-mechanical potential model described by Hamiltonian 
\begin{eqnarray}
H=H_0+V(r); \qquad H_0=\frac{k^2}{2m}, \qquad V(r)=V_{\rm conf}(r)-\frac{\alpha}{r}.
\end{eqnarray}
Polarization operator $\Pi(E)$ and its Borel transform $\Pi(T)$ [$E\to T$, $1/(H-E)\to \exp(-H T)$, $T$ the Borel parameter] 
are defined via the full Green function $G(E)=1/(H-E)$ \cite{nsvz}: 
\begin{eqnarray}
\Pi(E)=\langle \vec r_f=0|\frac{1}{H-E}|\vec r_i=0\rangle,\qquad 
\Pi(T)=\langle \vec r_f=0|\exp(- H T)|\vec r_i=0\rangle.  
\end{eqnarray}
The expansion of $G(E)$ and $\Pi(E)$ in powers of the interaction is obtained with the help of the Lippmann-Schwinger equation 
\begin{eqnarray}
\label{lipp}
G(E)=G_0(E)-G_0(E)V G_0(E)+G_0(E)V G_0(E)V G_0(E)+\ldots,
\end{eqnarray}
where $G_0(E)=1/(H_0-E)$. Since the interaction contains now two parts, $V_{\rm conf}(r)$ and $\frac{\alpha}{r}$, 
the expansion (\ref{lipp}) is a double expansion in powers of $V_{\rm conf}$ and $\alpha$. 
E.g., for the case $V_{\rm conf}(r)=\frac{m\omega^2 r^2}2$ one easily obtains the corresponding 
double expansion in powers of $\alpha$ and $\omega T$: 
\begin{eqnarray}
\Pi_{\rm OPE}(T)&=&\Pi_{\rm pert}(T)+\Pi_{\rm power}(T), \quad 
\Pi_{\rm pert}(T)=
\left(\frac{m}{2\pi T}\right)^{3/2}
\left[1+\sqrt{2\pi mT}\alpha+\frac{1}{3}m\pi^2 T \alpha^2\right], \quad 
\nonumber\\
\Pi_{\rm power}(T)&=&
\left(\frac{m}{2\pi T}\right)^{3/2}
\left[-\frac{1}{4}\omega^2 T^2\left(1+\frac{11}{12}\sqrt{2\pi m T}\alpha\right)
+\frac{19}{480}\omega^4 T^4\right].
\end{eqnarray}
One can see here a ``perturbative contribution'' (i.e. not containing the confining potential),
and ``power corrections'' given in terms of the confining potential (including also mixed 
terms containing contributions from both Coulomb and confining potentials). 
A perturbative contribution may be written in the form of spectral representation \cite{voloshin} yielding
\begin{eqnarray}
\Pi_{\rm OPE}(T)=\int\limits_0^\infty dz e^{- z T}\rho_{\rm pert}(z)+\Pi_{\rm power}(T),\;
\rho_{\rm pert}=
\left(\frac{m}{2\pi}\right)^{3/2}\left[2\sqrt{\frac{z}{\pi}}+\sqrt{2\pi m}\alpha+\frac{\pi^{3/2} m\alpha^2}{3\sqrt{z}}\right]
\end{eqnarray}
The ``physical'' representation for $\Pi(T)$---in the basis of hadron eigenstates---reads: 
\begin{eqnarray}
\Pi_{\rm phys}(T)=\langle \vec r_f=0|\exp(-H T)|\vec r_i=0\rangle = \sum_{n=0}^\infty R_n \exp(-E_n T), \quad R_n=|\Psi_n(\vec r=0)|^2.
\end{eqnarray}
Sum rule is the expression of the fact that the correlator may be calculated in two ways---using  
the basis of quark states (OPE) or confined bound states---leading to the same result: 
\begin{eqnarray}
\Pi_{\rm OPE}(T)=\Pi_{\rm phys}(T). 
\end{eqnarray}
In order to isolate the ground-state contribution one needs the information about the excited states. 
A standard Ansatz for the hadron spectral density has the form \cite{svz} 
\begin{eqnarray}
\label{ansatz}
\rho_{\rm phys}(z)=R_g\delta(z-E_g)+\theta(z-z_{\rm eff})\rho_{\rm pert}(z).
\end{eqnarray}
It assumes that the contribution of the excited states may be described by contributions of diagrams of 
perturbation theory above some effective continuum threshold $z_{\rm eff}$. This effective continuum threhsold  
(different from the physical continuum threhsold which is determined by hadron masses) is an additional 
parameter of the method of sum rules. Using Eq.~(\ref{ansatz}) yields 
\begin{eqnarray}
R_g e^{- E_g T}=\int\limits_0^{z_{\rm eff}}dz e^{- z T}\rho_{\rm pert}(z)+\Pi_{\rm power}(T)\equiv \Pi_{\rm dual}(T,z_{\rm eff}).
\end{eqnarray}
As soon as one knows $z_{\rm eff}$, one immediately obtains estimates for $R_g$ and $E_g$, 
$R_{\rm dual}(T,z_{\rm eff})$ and $E_{\rm dual}(T,z_{\rm eff})=-d_T \log \Pi(T,z_{\rm eff})$. 
These however depend on unphysical parameters $T$ and $z_{\rm eff}$. 

Eq.~(\ref{ansatz}) is motivated by quark-hadron duality which claims that far above 
the threshold the hadron spectral density is 
well described by diagrams of perturbation theory. However, near the physical threshold---and this very region turns 
out to be essentail for the calculation of ground-state properties---the duality relation is violated. 

The advantage of quantum mechanics is that the exact $E_g$ and $R_g$ may be obtained by solving Schr\"odinger 
equation \cite{schoberl}. We can then calculate $z_{\rm eff}$ from 
\begin{eqnarray}
R_g e^{- E_g T}=\int\limits_0^{z_{\rm eff}}dz e^{- z T}\rho_{\rm pert}(z)+\Pi_{\rm power}(T). 
\end{eqnarray}
The obtained ``exact threshold'' $z_{\rm eff}(T)$ is a slightly rising function of $T$ \cite{lms_2ptsr}. 

\section{OPE and sum rule in QCD} 
In QCD, the sum rule for the decay constants of heavy mesons has a similar form \cite{jamin}
\begin{eqnarray}
\label{SR_QCD}
f_Q^2 M_Q^4 e^{-M_Q^2\tau}=
\int\limits^{s_{\rm eff}}_{(m_Q+m_u)^2} e^{-s\tau}\rho_{\rm pert}(s,\alpha,m_Q,\mu)\,ds + 
\Pi_{\rm power}(\tau,m_Q,\mu)\equiv \Pi_{\rm dual}(\tau,\mu,s_{\rm eff})
\end{eqnarray}
In order to extract the decay constant one should fix the 
effective continuum threshold $s_{\rm eff}$ which should be a function of $\tau$;  
otherwise the $\tau$-dependences of the l.h.s. and the r.h.s. of (\ref{SR_QCD}) 
do not match each other. 
The exact $s_{\rm eff}$ corresponding to the exact hadron decay constant 
and mass on the l.h.s. is of course not known. The extraction of hadron parameters from the 
sum rule consists therefore in attempting 
(i) to find a good approximation to the exact threshold and 
(ii) to control the accuracy of this approximation. 
For further use, we define the dual decay constant $f_{\rm dual}$ and the dual invariant mass $M_{\rm dual}$ by relations
\begin{eqnarray}
\label{fdual}
f_{\rm dual}^2(\tau)=M_Q^{-4} e^{M_Q^2\tau}\Pi_{\rm dual}(\tau, s_{\rm eff}(\tau)),\quad 
M_{\rm dual}^2(\tau)=-\frac{d}{d\tau}\log \Pi_{\rm dual}(\tau, s_{\rm eff}(\tau)). 
\end{eqnarray}

\section{Extraction of the decay constant: QCD vs. potential model \cite{lms_qcdvsqm}}
If the mass of the ground state is known, the deviation of the dual mass from the actual mass of the ground state 
gives an indication of the contamination of the dual correlator by excited states. 
Assuming a specific functional form of the effective threshold and requiring the least deviation of the dual mass 
(\ref{fdual}) from the known ground-state mass in the $\tau$-window leads to a variational solution for the 
effective threshold. As soon as the latter has been fixed, one calculates the decay constant from (\ref{fdual}). 
{\bf Our algorithm for extracting ground-state parameters reads:}

\noindent (i) Consider a set of Polynomial $\tau$-dependent Ansaetze for $s_{\rm eff}$: 
$s^{(n)}_{\rm eff}(\tau)=s^{(n)}_0+s^{(n)}_1\tau+s^{(n)}_2 \tau^2+\ldots.$ 

\noindent (ii) Calculate the dual mass for these $\tau$-dependent thresholds and minimize the squared difference between the ``dual'' mass  
$M^2_{\rm dual}$ and the known value $M^2_B$ in the $\tau$-window.  
This gives the parameters of the effective continuum thresholds $s^{(n)}_i$. 

\noindent (iii) Making use of the obtained thresholds, calculate the decay constant. 

\noindent (iv) Take the band of values provided by the results corresponding to {\bf linear}, 
{\bf quadratic}, and {\bf cubic}
effective thresholds as the characteristic of the intrinsic uncertainty of the extraction procedure. 

\begin{figure}[!hb]
\begin{center}
\begin{tabular}{lr}
\includegraphics[width=6cm]{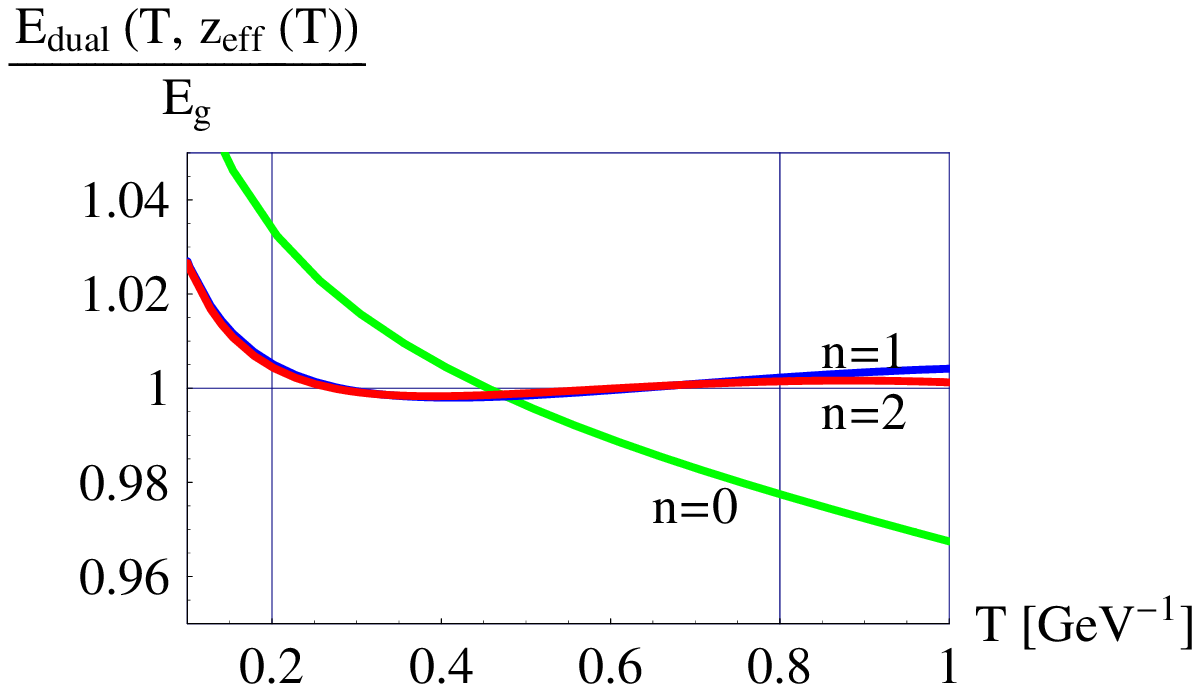}&\includegraphics[width=6cm]{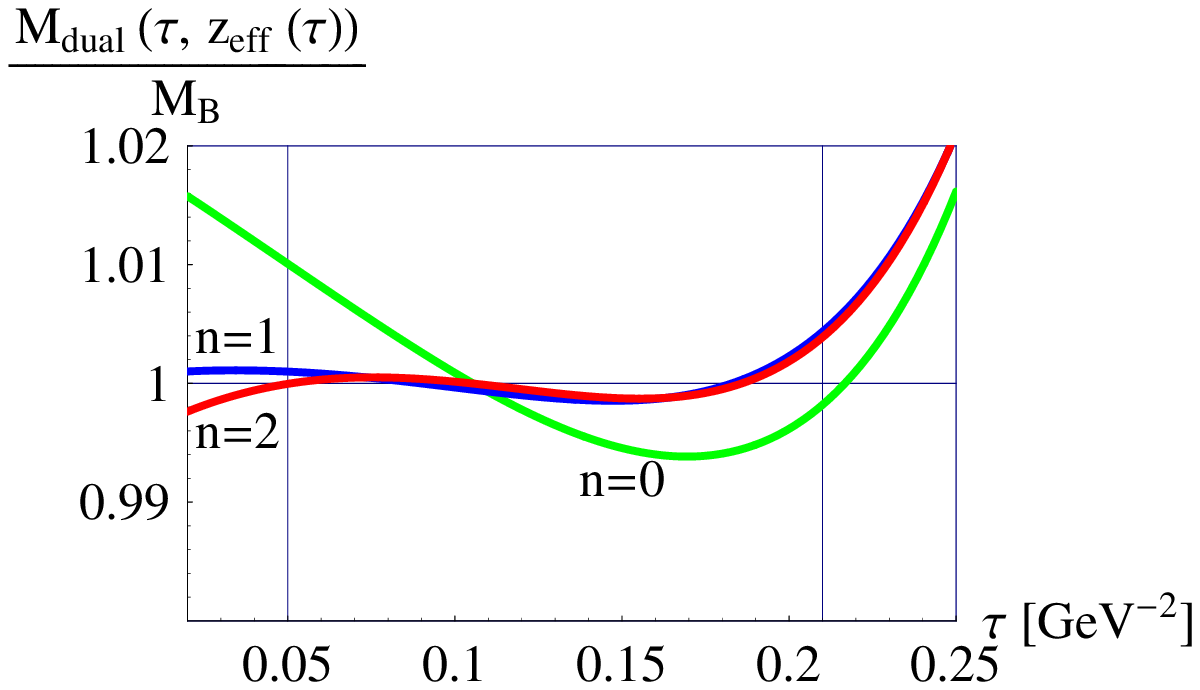}\\
\includegraphics[width=6cm]{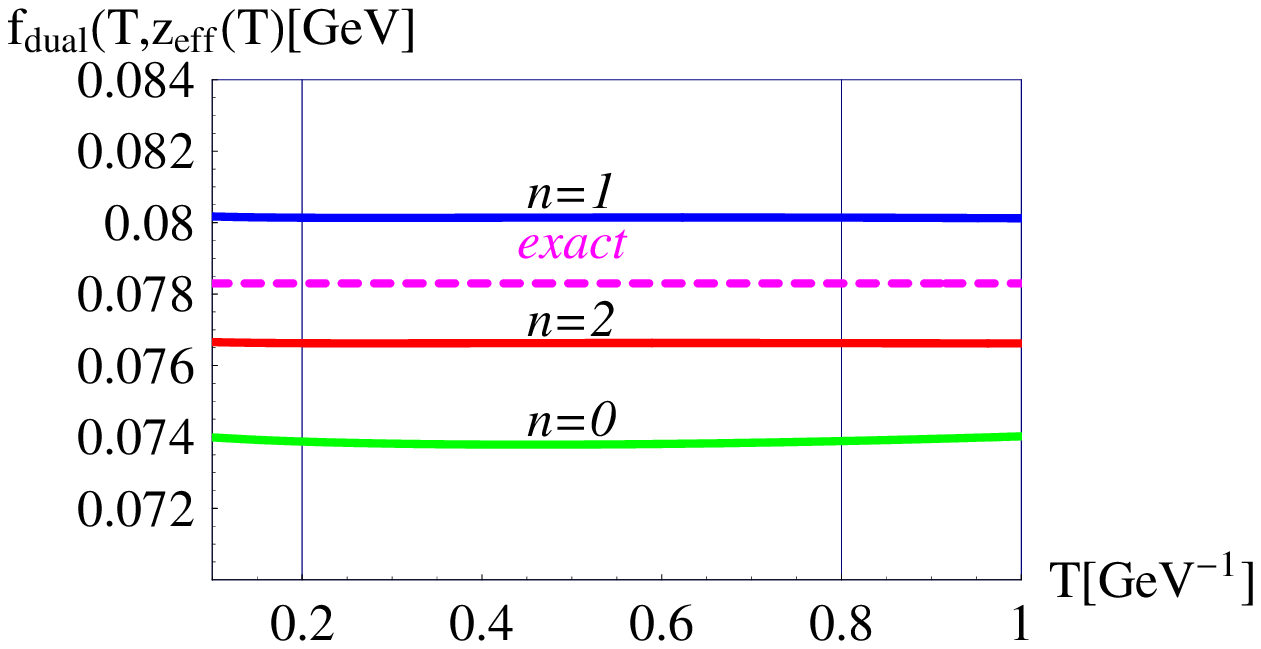}&\includegraphics[width=6cm]{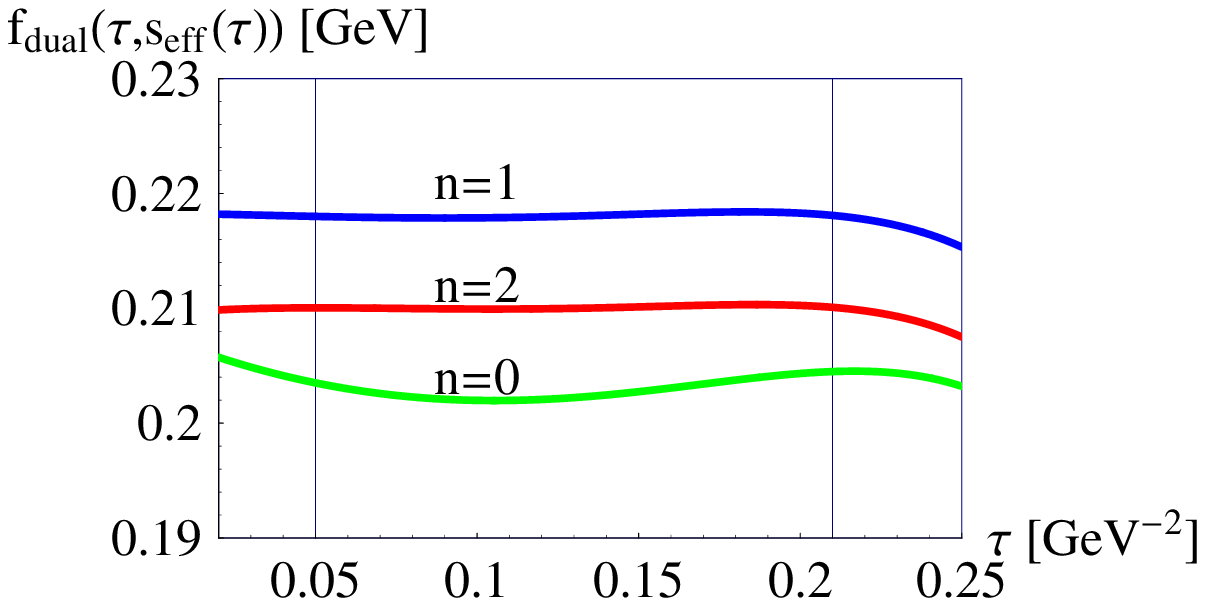}
\end{tabular}
\caption{\label{Fig:1}
The outcome of our algorithm in potential model (left) and in QCD (right).}
\end{center}
\end{figure}

Figure \ref{Fig:1} shows the application of our algorithm in quantum mechanics and in QCD. 
In both cases a very similar pattern emerges; 
the reproduction of the dual mass considerably improves for the $\tau$-dependent quantities indicating that the dual 
correlator with a $\tau$-dependent threshold isolates the contribution of the ground state much better 
than the dual correlator with a standard $\tau$-independent threshold.
As a consequence, the accuracy of the extracted hadron observable improves considerably. 

\vspace{-.4cm}
\section{Decay constants of $D$ and $D_s$ mesons \cite{lms_fp}}
\begin{figure}[!ht]
\begin{center}
\begin{tabular}{cc}
\includegraphics[height=5.5cm]{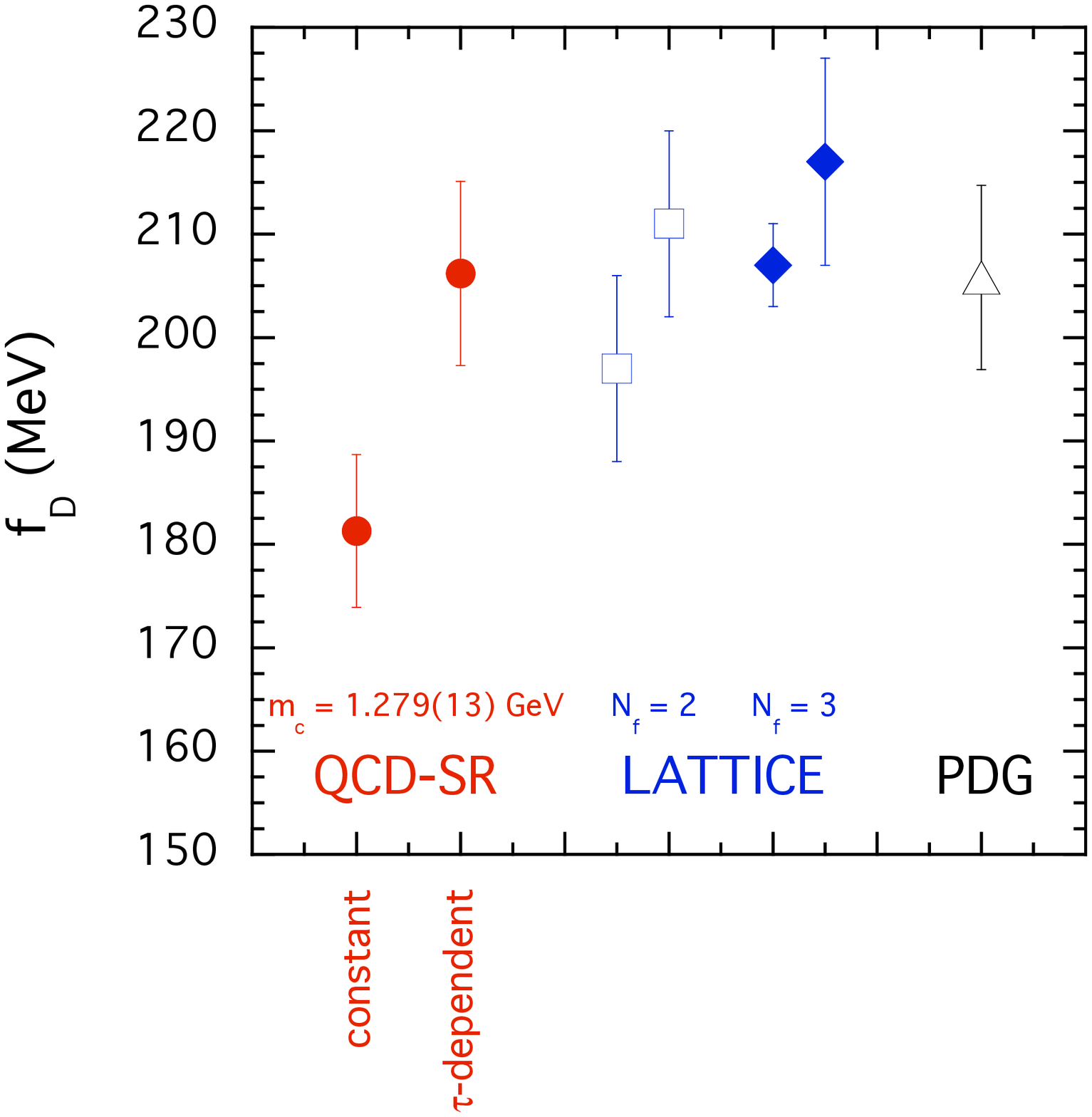}&
\includegraphics[height=5.5cm]{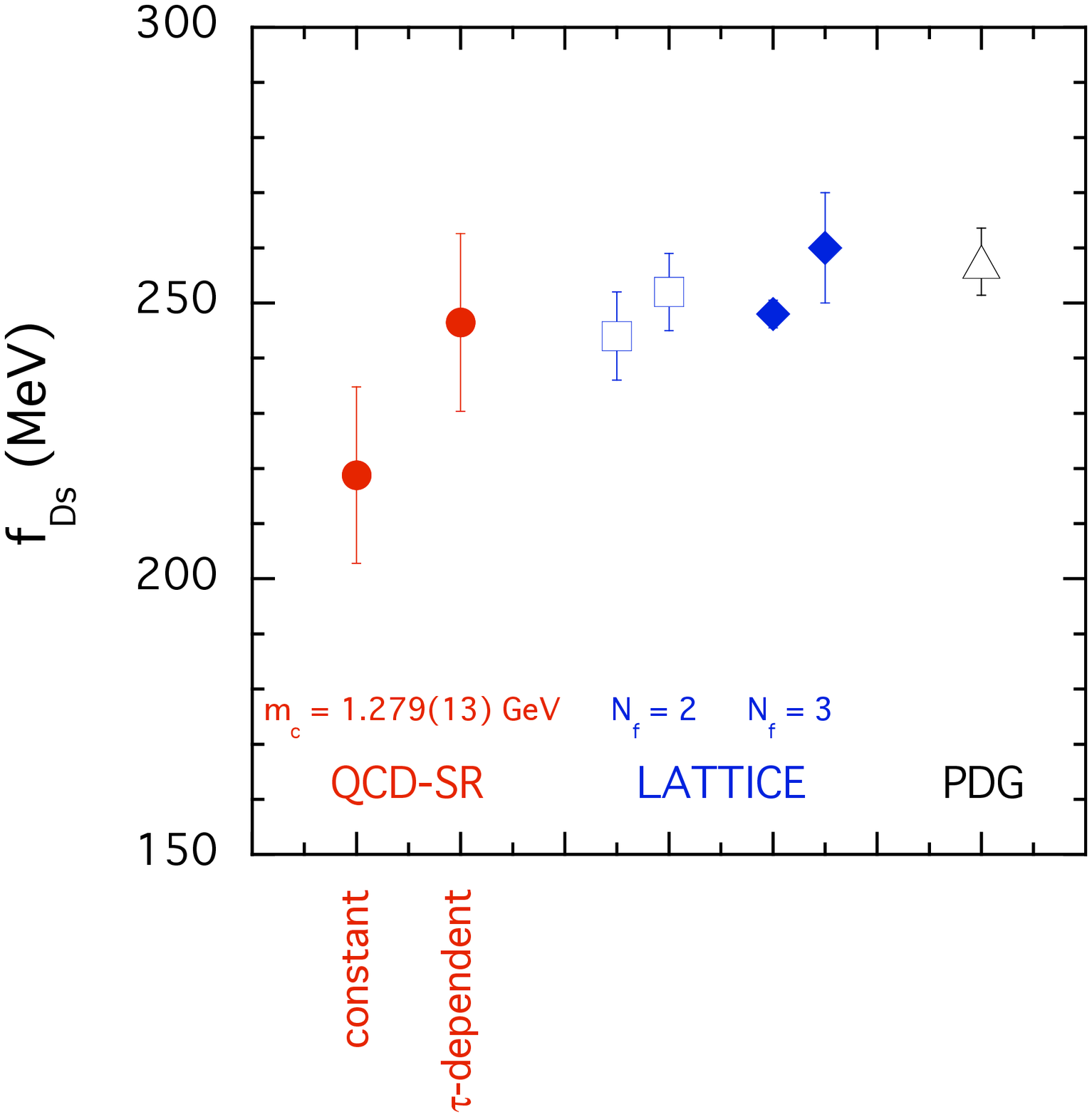}
\end{tabular}
\caption{\label{Plot:Dresults}
Our results for $f_D$ and $f_{Ds}$ \cite{lms_fp}.}
\end{center}
\end{figure}

The application of our extraction procedures leads to the following results:
\begin{eqnarray}
\label{Dresults}
f_{D}= 206.2 \pm 7.3_{\rm (OPE)} \pm 5.1_{\rm (syst)}  \mbox{MeV},\qquad 
f_{D_s}= 245.3 \pm 15.7_{\rm (OPE)} \pm 4.5_{\rm (syst)}\; {\rm MeV}.   
\end{eqnarray}
The $\tau$-dependent threshold is a crucial ingredient for a successful extraction 
of the decay constant from the sum rule (Fig.~\ref{Plot:Dresults}). 
Obviously, the standard $\tau$-independent approximation leads to much lower 
value for $f_D$ which lies rather far from the data and the lattice results. 
\newpage

${}$
\vspace{-1.5cm}
\section{Conclusions}
The effective continuum threshold $s_{\rm eff}$ is an important ingredient of the method of dispersive 
sum rules which determines to a large extent the numerical values of the extracted hadron 
parameter. Finding a criterion for fixing $s_{\rm eff}$ poses a problem in the method of sum rules. 

\vspace{.1cm}
\noindent 
$\bullet$ $s_{\rm eff}$ depends on the {\it external kinematical variables} (e.g., 
momentum transfer in sum rules for 3-point correlators and light-cone sum rules) and 
{`\it `unphysical'' parameters} (renormalization scale $\mu$, Borel parameter $\tau$). 
Borel-parameter $\tau$-dependence of $s_{\rm eff}$ emerges naturally when trying to make 
quark-hadron duality more accurate. 

\vspace{.1cm}\noindent 
$\bullet$ We proposed a new algorithm for fixing $\tau$-dependent $s_{\rm eff}$, for those problems 
where the bound-state mass is known. We have tested that our algorithm leads to more accurate 
values of ground-state parameters than the ``standard'' algorithms used in the context 
of dispersive sum rules before. Moreover, our algorithm allows one to probe 
{\it ``intrinsic''} uncertainties related to the limited accuracy of 
the extraction procedure in the method of QCD sum rules.  
\vspace{.3cm}

\noindent 
{\it Acknowledgments.} 
I have pleasure to thank Wolfgang Lucha and Silvano Simula for the most pleasant and fruitful collaboration 
on the subject presented in this lecture and to the Austrian Science Fund FWF for support under Project P22843.
Special thanks are due to the Organizers for creating a stimulating and pleasant atmosphere of this Conference.

\end{document}